\documentclass{article}
\usepackage{spconf,amsmath,graphicx}
\usepackage{multirow}
\usepackage{subcaption}
\usepackage{caption}

\title{Domain Expansion in DNN-based Acoustic Models for\\Robust Speech Recognition}
\name{Shahram Ghorbani, Soheil Khorram, John H.L. Hansen\thanks{This project was supported in part by AFRL under contract FA8750-15-1-0205, and partially by the University of Texas at Dallas from the Distinguished University Chair in Telecommunications Engineering held by J. H. L. Hansen.}}
\address{Center for Robust Speech Systems (CRSS), 
University of Texas at Dallas}

\begin{document}
%
%
%
%
\maketitle
\begin{abstract}
Training acoustic models with sequentially incoming data -- \emph{while both leveraging new data and avoiding the forgetting effect} -- is an essential obstacle to achieving human intelligence level in speech recognition. An obvious approach to leverage data from a new domain (e.g., new accented speech) is to first generate a comprehensive dataset of all domains, by combining all available data, and then use this dataset to retrain the acoustic models. However, as the amount of training data grows, storing and retraining on such a large-scale dataset becomes practically impossible. To deal with this problem, in this study, we study several domain expansion techniques which exploit only the data of the new domain to build a stronger model for all domains. These techniques are aimed at learning the new domain with a minimal forgetting effect (i.e., they maintain original model performance). These techniques modify the adaptation procedure by imposing new constraints including (1) \emph{weight constraint adaptation (WCA)}: keeping the model parameters close to the original model parameters; (2) \emph{elastic weight consolidation (EWC)}: slowing down training for parameters that are important for previously established domains; (3) \emph{soft KL-divergence (SKLD)}: restricting the KL-divergence between the original and the adapted model output distributions; and (4) \emph{hybrid SKLD-EWC}: incorporating both SKLD and EWC constraints. We evaluate these techniques in an accent adaptation task in which we adapt a deep neural network (DNN) acoustic model trained with native English to three different English accents: Australian, Hispanic, and Indian. The experimental results show that SKLD significantly outperforms EWC, and EWC works better than WCA. The hybrid SKLD-EWC technique results in the best overall performance.

%
\end{abstract}
\section{Introduction}
\label{sec:intro}

Current state-of-the-art neural network-based ASR systems have advanced to nearly human performance in several evaluation settings~\cite{saon2017english, best2}; however, these systems perform poorly for domains\footnote{In this paper, we use the term "domain" to refer to a group of utterances that share some common characteristics.} that are not included in the original training data~\cite{myself1,sun2017unsupervised,hsu2017unsupervised,jafarlou2019LRF}. For example, if we train an ASR system using a U.S. English dataset, the performance of the system significantly degrades for other English accents (e.g., Australian, Indian, and Hispanic). In order to improve performance of the system for an unseen domain, we can adapt the previously trained model to capture the statistics of the new domain. However, adaptation techniques suffer from the \emph{forgetting effect}: previously learned information will be lost by learning the new information. We need an ASR system that not only performs well for the new domain, but also retains performance for previously seen domains. This is the goal of \emph{domain expansion} methods.
 
\textbf{Domain Expansion} -- In a domain expansion scenario, we are given a model trained on an initial domain and a dataset for an unseen domain, the goal is to modify the model such that it performs well for both domains. The main difficulty of domain expansion is to preserve the functionality (input-output mapping) of the original model (mitigating the forgetting problem). Many approaches have been proposed to deal with the forgetting problem in neural networks. These approaches can be divided into three categories: architectural, rehearsal, and regularization strategies. 

\subsection{Architectural strategies}
In this class of methods, architectures of neural networks are modified to mitigate the forgetting problem. Progressive neural network (PNN)~\cite{rusu2016progressive} is a popular architectural strategy; it freezes the previously trained network and uses its intermediate representations as inputs into a new smaller network. PNN has been applied in many different applications including speech synthesis~\cite{hodari2018learning}, speaker identification~\cite{gideon2017progressive} and speech emotion recognition~\cite{gideon2017progressive}. 
However, it has been shown that PNN is not efficient for long sequences of domains, since the number of weights in PNN increases linearly with the number of domains~\cite{rusu2016progressive}. 

\subsection{Rehearsal strategies}
These approaches store part of the previous training data and periodically replay them for future training. A full rehearsal strategy can alleviate the forgetting effect, but it is very slow and memory intensive. Tylor et al. proposed EXSTREAM, a new partitioning-based approach, to address the memory problem of the full rehearsal strategy~\cite{hayes2018memory}. In another approach, \cite{draelos2017neurogenesis} proposed to train an encoder-decoder model that distills information which exists in the previous domains. Their method uses the trained encoder-decoder to simulate pseudo patterns of the previous domains and exploits these pseudo patterns during the training of the new domain. 

\subsection{Regularization strategies}
Regularization refers to a set of techniques that alleviate the forgetting effect by imposing additional constraints on updating parameters. A straightforward constraint is \emph{weight constraint adaptation (WCA)} which penalizes the deviation of the model parameters from the original model parameters; it adds an $l_2$ distance between the original and adapted weights~\cite{li2006regularized}. Another popular regularization approach is \emph{learning without forgetting (LWF)}~\cite{lwf} that tries to learn a sequence of relevant tasks without losing performance for the older ones by imposing output stability. Jung et al.~\cite{jung2018less} explored the domain expansion problem for image classification tasks. They used an $l_2$ distance between the final hidden representations of the original network and the adapted network. Kirkpatrick et al.~\cite{EWC} introduced \emph{elastic weight consolidation (EWC)} which selectively slows down the training for weights that are important for older domains. 

In this study, we explore approaches to address the domain expansion problem for the deep neural network (DNN)-based acoustic models. To the best of our knowledge, this is the first study that explores domain expansion for speech recognition. We investigate several existing and proposed regularization strategies to alleviate the forgetting effect in domain expansion. We employ WCA and EWC as the baseline techniques for the domain expansion problem; we also propose two new domain expansion techniques: \emph{soft KL-divergence (SKLD)} and hybrid SKLD-EWC. SKLD penalizes the KL-divergence (KLD) between the original model's output and the adapted model's output as a measure of the deviation of the model. We will demonstrate that the proposed SKLD and EWC are complementary to each other, and combining them can lead to a better domain expansion technique which we refer to as SKLD-EWC. We will compare the efficacy of these methods in an accent adaptation task in which we adapt a DNN acoustic model trained with native English to three different English accents: Australian, Hispanic, and Indian. Our results will show that the proposed hybrid technique, SKLD-EWC, results in the best overall performance and SKLD performs significantly better than EWC and WCA. 

\section{Domain Expansion Approaches}

In this section, we explain details of four domain expansion techniques (i.e., weight constraint adaptation (WCA), elastic weight consolidation (EWC), soft KL-Divergence (SKLD), and hybrid SKLD-EWC) that we investigate in this study.

\textbf{Problem Setup} --
In the domain expansion task, we are given an original model $\mathcal{M}^o$, trained on an original domain $\mathcal{D}^o$, and a dataset for an unseen domain $\mathcal{D}^n$, where the goal is to find a new model $\mathcal{M}^n$ that performs well for both $\mathcal{D}^o$ and $\mathcal{D}^n$.

\subsection{Weight Constraint Adaptation (WCA)}

WCA was first proposed in~\cite{li2006regularized} to regularize the adaptation process for discriminative classifiers. In another study~\cite{lwf}, WCA was employed for continual learning in a sequence of disjoint tasks. This technique tries to find a solution that performs well for the new domain, $\mathcal{D}^n$, which is also close to the original model, $\mathcal{M}^o$. 

According to~\cite{EWC}, for a given neural network architecture, there are many configurations of model parameters that lead to comparable performance. Therefore, there are many configurations that can efficiently represent our new domain $\mathcal{D}^n$. Among such configurations, an effective solution for domain expansion is the one that stands closer to the original model $\mathcal{M}^o$. Different distance metrics can be used to measure the similarity between models. WCA benefits from the Euclidean distance between the learnable parameters of $\mathcal{D}^o$ and $\mathcal{D}^n$. This idea can be implemented by imposing an additional $L_2$ constraint on the optimization loss function of the neural network:
\begin{equation} \label{eq:11}
    J_{WCA}(\theta^n) =  J_{cross}(\theta^n) +  \frac{\lambda_w}{2}  ||\theta^n - \theta^o||_2,
\end{equation}
where $\theta^o$ and $\theta^n$ are the learnable parameters of $\mathcal{M}^o$ and $\mathcal{M}^n$, respectively; $J_{cross}(\theta^n)$ is the main optimization loss (cross-entropy loss function); $J_{WCA}(\theta^n)$ is the regularized loss with the WCA technique; $||.||_2$ is the $L_2$ norm; and $\lambda_w$ is a regularization parameter that determines how far the parameters could diverge from their initial values to learn the new domain.

\subsection{Elastic Weight Consolidation (EWC)}

The WCA technique considers all weights equally. Therefore, it is unable to find an efficient compromise to maintain the model performance for the original domain $\mathcal{D}^o$ and learning the new domain $\mathcal{D}^n$. However, all weights are not equally important, and using an approach that takes weight importance into account would perform better than a naive WCA.
 
Intuitively, after training a DNN with sufficient iterations, the model converges to a local minimum point of the optimization landscape. At such a point, the sensitivity of the loss function w.r.t. the $i$-th learnable weight, $\theta^n_i$, can be calculated by the curvature of the loss function along the direction specified by $\theta^n_i$ changes. High curvature for a weight means that the loss function is sensitive to small changes to that weight. Therefore, to preserve the performance of the network for the previous domain, we must prevent modifying the parameters with high curvature. On the other hand, parameters with low curvature values are proper choices to be tuned with new data without losing the model performance for the original data. 

The curvature of the loss function is equivalent to the diagonal of the Fisher information matrix $F$~\cite{maltoni2018continuous}. EWC offers a straightforward method to incorporate the importance of the learnable weights (curvature of the loss function w.r.t. the weights) in the adaptation process. The method is similar to WCA; the only difference is that EWC employs a weighted $L_2$ norm instead of the regular $L_2$ norm in WCA:
\begin{equation} \label{eq:1}
    J_{EWC}(\theta^n) =  J_{cross}(\theta^n) +  \frac{\lambda_e}{2} \sum_{i} diag\{F\}_i (\theta^n_i - \theta^o_i)^2,
\end{equation}
where $diag\{F\}_i$ is the $i$-th element of the diagonal of the Fisher information matrix $F$ (representing the importance of the $i$-th learnable weight); $\theta^n_i$ and $\theta^o_i$ are the $i$-th weight of the new and original models, respectively; and the summation is taken over all learnable weights of the network. $diag\{F\}$ can be easily calculated by the variance of the first order derivatives of the loss function w.r.t. the learnable weights (i.e., $Var\{\partial J(\theta)/\partial \theta_i$\})~\cite{maltoni2018continuous}.

\subsection{Soft KL-Divergence (SKLD)}
A major difficulty of the domain expansion task is to preserve the functionality (input-output mapping) of the original model $\mathcal{M}^o$. WCA and EWC achieve this by providing a link between the learnable weights of the new model $\mathcal{M}^n$ and the original model $\mathcal{M}^o$. According to the experiments performed in ~\cite{jung2018less}, linking the learnable parameters is not an efficient way of preserving the functionality of the parameters, since applying slight changes to some of the parameters may significantly modify the input-output mapping of the network. 
Another method for preserving the functionality of $\mathcal{M}^o$ is to impose new constraints on the outputs of the model \cite{yu2013kl}. By constraining the outputs of $\mathcal{M}^n$ to be consistent with the outputs of $\mathcal{M}^o$, we can assure that these two models are similar to each other. SKLD leverages this idea through two steps: (1) it takes the original model $\mathcal{M}^o$ and the data of the new domain $\mathcal{D}^n$; it then generates the output of $\mathcal{M}^o$ for all samples of the dataset. (2) next, SKLD trains the new model $\mathcal{M}^n$ by initializing it from $\mathcal{M}^o$ and using a regularized loss function that can be expressed by:
\begin{multline}\label{eq:SKLD1}
    J_{SKLD}(\theta^n) =  (1-\lambda_s) J_{cross}(\theta^n) + \\
    \lambda_s \sum_{i \in I} D_{KL}(\mathcal{M}^n(x_i), \mathcal{M}^o(x_i)),
\end{multline}
where $I$ is the total number of samples; $D_{KL}$ is the KL distance; $x_i$ is the $i$-th input feature vector; $\mathcal{M}^o(x_i)$ and $\mathcal{M}^n(x_i)$ are the outputs of the original and the new models obtained for the $i$-th sample $x_i$; and $0 \le \lambda_s \le 1$ is a regularization hyper-parameter that provides a compromise between learning the new domain (by optimizing $J_{cross}$) and preserving the input-output mapping of the original model (by optimizing $D_{KL}$). $\lambda_s = 0$ results in the conventional pre-training/fine-tuning adaptation. By increasing the value of $\lambda_s$, we can ensure a balanced trade-off between learning the new domain and mitigating the forgetting effect problem. For this study, we tune $\lambda_s$ to achieve the best performance for both domains. 

Equation (\ref{eq:SKLD1}) uses the KL divergence between $\mathcal{M}^o(x_i)$ and $\mathcal{M}^n(x_i)$ to deal with the forgetting problem. However, some parts of the KL divergence are not related to the learnable parameters of $\mathcal{M}^n$. In \cite{kldr1}, it is demonstrated that by removing these parts, the KL divergence will be simplified to the cross-entropy:
\begin{multline} \label{eq:SKLD2}
    J_{SKLD}(\theta^n) =  (1-\lambda_s) J_{cross}(\theta^n) + \\ 
    \hspace{3.5cm} \lambda_s \sum_{i \in I} J_{cross}(\mathcal{M}^o(x_i), \mathcal{M}^n(x_i)),\\
    J_{cross}(\mathcal{M}^o(x_i), \mathcal{M}^n(x_i)) = \hspace{3.5cm} \\ 
    \sum_{c \in C} \mathcal{M}_c^o(x_i) \log(\mathcal{M}_c^n(x_i)),
\end{multline}
where $C$ is the total number of classes; $\mathcal{M}_c^o(x_i)$ and $\mathcal{M}_c^n(x_i)$ are the probability of the $c$-th class generated by $\mathcal{M}^o$ and $\mathcal{M}^n$ for an input vector $x_i$. 

In neural networks, we typically use a softmax with temperature $T = 1$ to produce the probability for each class. However, Hinton et al. \cite{hinton2015distilling} suggested that using $T > 1$ that increases the probability of small logits, performs better in transferring the functionality of a large network to a smaller one. Therefore, we also consider tuning the temperature to examine its effects in preserving the functionality of the model for the original data. We consider using a softmax with adjustable temperature to produce the output distribution for both $\mathcal{M}^o$ and $\mathcal{M}^n$ in the constraint term of equations (\ref{eq:SKLD2}). Note that we use $T = 1$ for the tuning loss as well as in the evaluation phase.

\subsection{Hybrid SKLD-EWC}

In the previous sections, we explained both SKLD and EWC approaches. Each one has its advantages and disadvantages. The advantage of EWC is that it computes the Fisher information matrix (that quantifies the importance of the weights) based on the original data during the initial training. However, SKLD does not exploit such information about the importance of the weights. On the other hand, EWC uses a fixed fisher matrix that is estimated for the initial model. However, fisher matrix changes during the adaptation procedure and therefore the fixed assumption of the EWC method is not reliable. The advantage of SKLD is that it is more efficient in preserving the functionality of the original model as the efficacy of SKLD does not change during the adaptation procedure. 
We propose to combine these two techniques into a new hybrid approach SKLD-EWC. Our proposed technique can be implemented by imposing both SKLD and EWC constraints on the tuning loss: 
\begin{multline} \label{eq:SKLD3}
    J_{SKLD-EWC}(\theta^n) = (1-\lambda_s) J_{cross}(\theta^n) + \\[4pt]
    \lambda_s \sum_{i \in I} J_{cross}(\mathcal{M}^o(x_i), \mathcal{M}^n(x_i)) + \\[-5pt]
    \lambda_e \sum_{i} diag\{F\}_i (\theta^n_i - \theta^o_i)^2. 
\end{multline}
This hybrid method requires two regularization parameters: $\lambda_s$ and $\lambda_e$ defined for regularizing the outputs and the weights, respectively. These two parameters provide a more flexible data expansion technique, but at the expense of more difficult hyper-parameter tuning.

\section{Experiments and Results}

\subsection{Experimental Setup}

\textbf{Dataset} -- We evaluate the efficacy of the domain expansion techniques for a DNN-HMM-based ASR system. To train the original model for native English, we use the 100h part of the LIBRISPEECH corpus (Libri) \cite{libri}. This part has higher recording quality and the speakers' accents are closer to the native English compared to the rest of the corpus. For the domain expansion experiments, we use UT-CRSS-4EnglishAccent corpus \cite{myself} that contains speech data from 3 non-US English accents, namely Hispanic (HIS), Indian (IND) and Australian (AUS). The data for each accent consists of 100 speakers, with session content that consists of read and spontaneous speech. In this corpus, for each accent, there is more than 28h of training data, 5h of development and 5h of evaluation data. We use the standard language model (LM) provided for Libri to decode the original data (i.e., Libri) \cite{libri}. However, for the other accents, since we have spontaneous utterances too, we train a 3-gram and a 4-gram LM by pooling transcriptions of Fisher, Switchboard, and UT-CRSS-4EnglishAccent. The decoding procedure used for Libri and other accents is the same.
\begin{table}  [ht]
\caption{WERs of different  domain expansion methods on the original (Org) and the new domains (New). For each approach, among the different settings of regularization parameters, the one that results in the best overall performance for both original and new domains (i.e., the lowest average of WERs) is reported. The relative WER increase of the methods compared to multi-condition (MC) training is also reported (Rel-MC)}
\label{table:1}
\begin{center}
\begin{tabular}{cccccc}
\hline
        &Method & Org &  New  & Avg&Rel-MC \\
      \hline
    \hline
    
    \multirow{7}{*}{\textbf{AUS}}
    &MC&8.35&10.7&9.52&---\\
    &Original&8.10&23.04&15.57&63.5\\
    & Fine-Tuned&20.3&9.64&14.97&57.2\\
    &WCA&10.23&12.2&11.22&17.74\\
    &EWC&9.27&12.71&11&15.38\\
    &SKLD&8.84&11.23&\textbf{10.03}&\textbf{5.35}\\
    &SKLD-EWC&8.49&11.48&\textbf{9.98}&\textbf{4.8}\\
    \hline
    \multirow{7}{*}{\textbf{IND}}
    &MC&8.29&15.96&12.12&---\\
    &Original&8.10&28.68&18.39&51.6\\
    & Fine-Tuned&22.54&15.56&19.05&57.1\\
    &WCA&10.66&17.61&14.13&16.5\\
    &EWC&10.26&17.6&13.93&14.8\\
    &SKLD&9.29&16.45&\textbf{12.87}&\textbf{6.14}\\
    &SKLD-EWC&8.92&16.61&\textbf{12.76}&\textbf{5.2}\\
    \hline
    \multirow{7}{*}{\textbf{HIS}}
     &MC&8.21&11.65&9.93&---\\
    &Original&8.10&20.09&14.1&41.9\\
    &  Fine-Tuned&16.14&11.52&13.83&39\\
    &WCA&8.65&12.84&10.74&8.2\\
    &EWC&8.72&12.61&10.66&7.4\\
    &SKLD&9.26&12.0&10.63&7.0\\
    &SKLD-EWC&8.29&12.5&\textbf{10.39}&\textbf{4.7}\\
    \hline
\end{tabular}
\vspace{-2em}
\end{center}
\label{tab:multicol}
\end{table}

\begin{figure*}[t]
        \begin{subfigure}[b]{0.32\textwidth}
                \includegraphics[width=\linewidth]{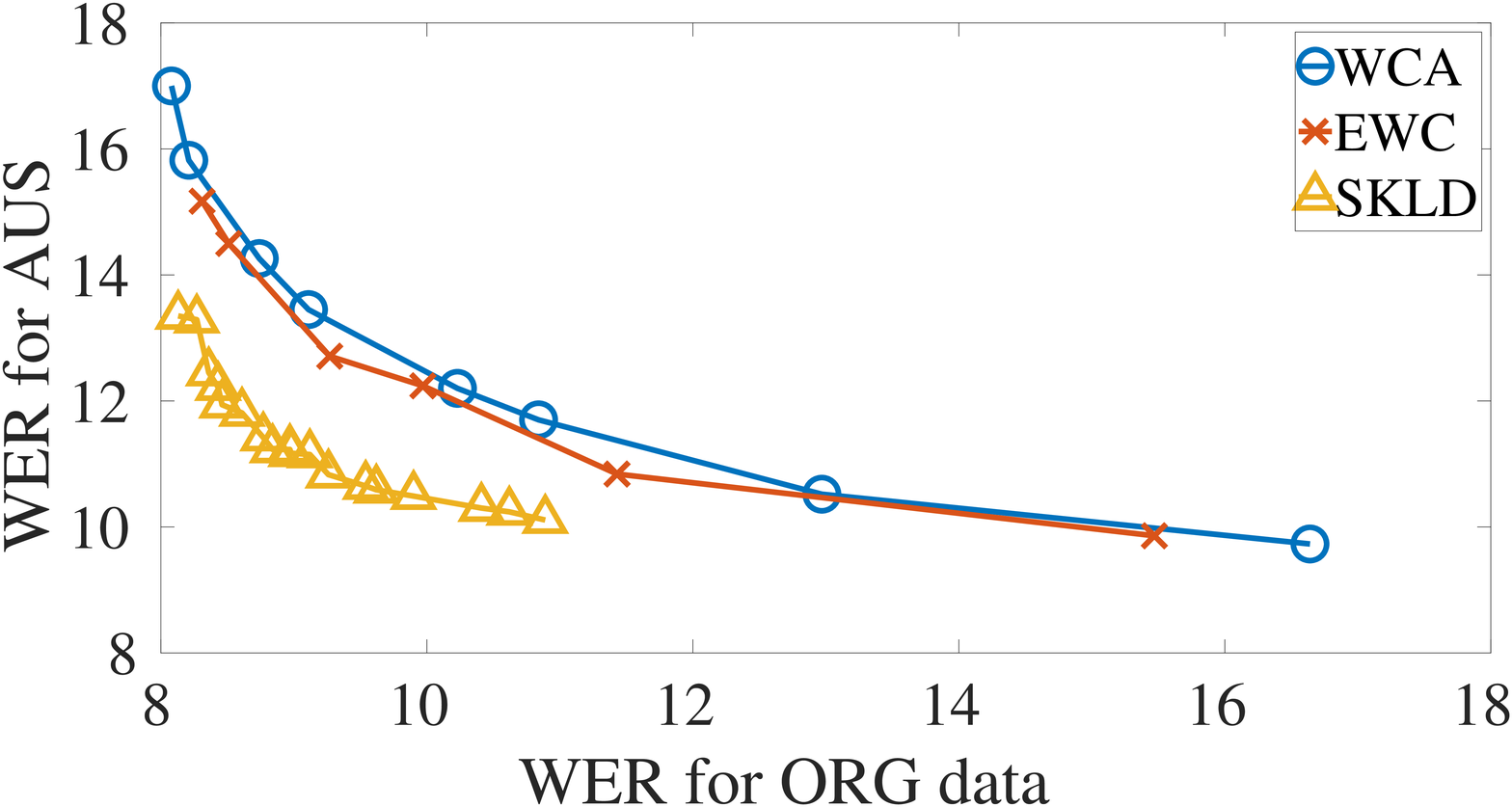}
                \label{fig:gull}
        \end{subfigure}\hspace{0.015\textwidth}
        \begin{subfigure}[b]{0.32\textwidth}
                \includegraphics[width=\linewidth]{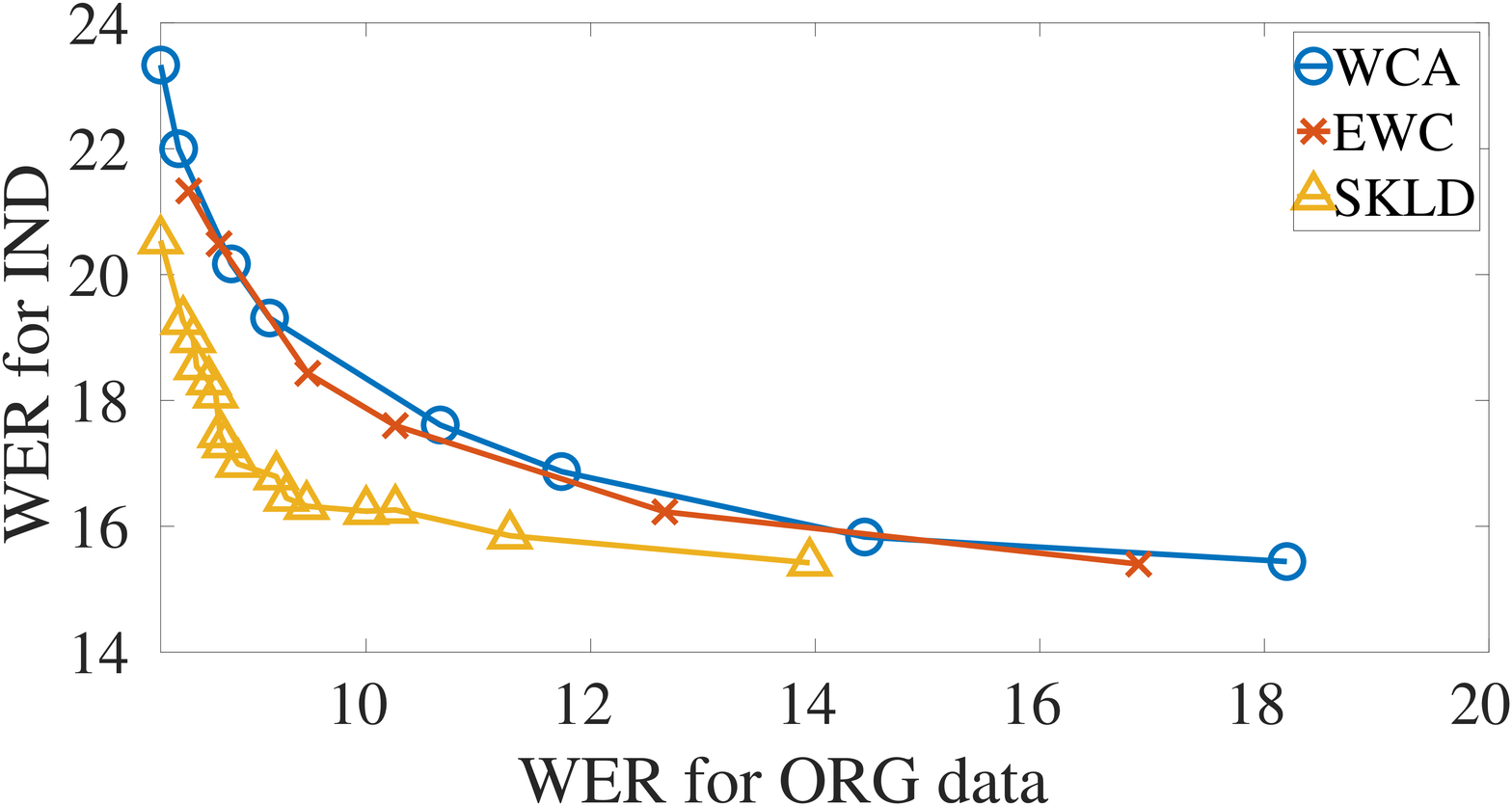}
                \label{fig:gull2}
        \end{subfigure}\hspace{0.015\textwidth}
        \begin{subfigure}[b]{0.32\textwidth}
                \includegraphics[width=\linewidth]{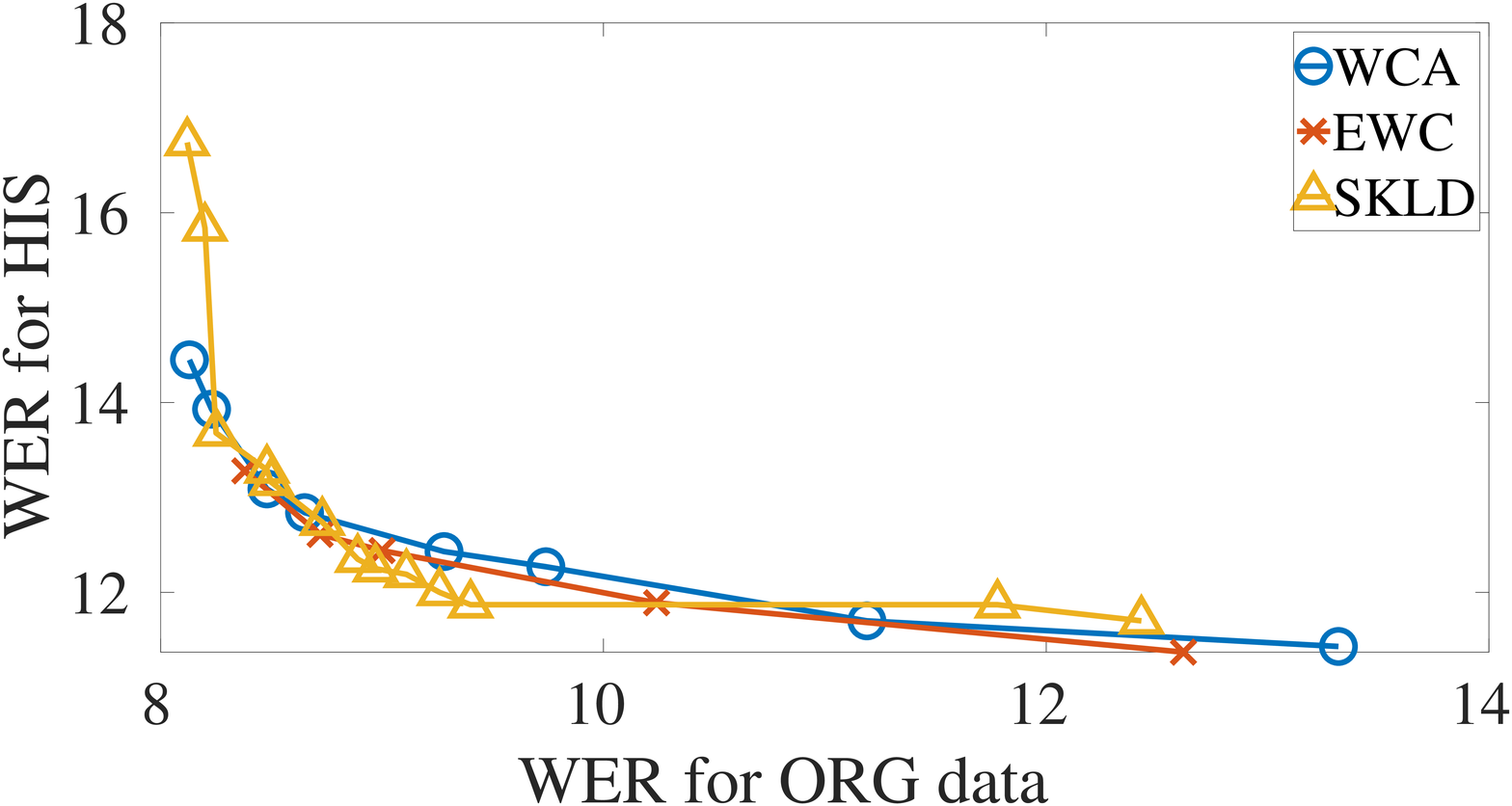}
                \label{fig:tiger}
        \end{subfigure}%
        \vspace{-10pt}
        \caption{Visualizing WER of different techniques on original and new datasets. These curves are generated by changing the hyper-parameters that control the trade-off between the performance of the original and the new domains.}\label{fig:compare}
\end{figure*}

\begin{figure}[t]
  \centering
  \includegraphics[width=8cm]{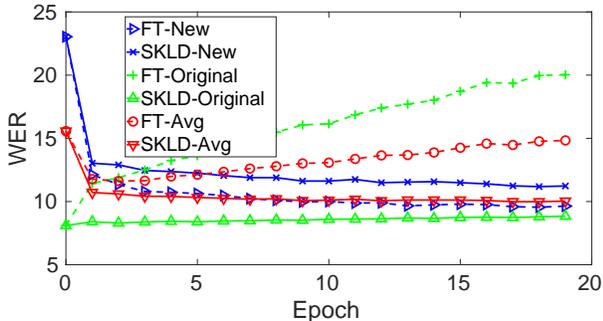}
  \vspace{-1em}
  \caption{Visualizing the forgetting effect of fine-tuning the original model to the AUS accent. The performance of SKLD in the same setting is also reported, which demonstrates the efficacy of SKLD to preserve the learned knowledge while adapting to the new domain. FT-Original: performance of the fine-tuned model on Libri; FT-New: performance of the fine-tuned model on AUS; FT-Avg: average of FT-Original and FT-new; SKLD-Original: performance of SKLD on Libri; SKLD-New: performance of SKLD on AUS; SKLD-Avg: average of SKLD-Original and SKLD-New}
  \label{fig:Forgetting}
\end{figure}

\textbf{Model structure} -- We implement the domain expansion techniques for a DNN-HMM based ASR system using Kaldi~\cite{povey2011kaldi} and Tensorflow~\cite{abadi2016tensorflow}. In all experiments, we extract 40-dimensional Mel-filterbank coefficients~\cite{povey2011kaldi} for each 25ms frame with a skip rate of 10ms. Each frame is expanded by stacking 5 frames from each side; therefore, the input to the network is the Mel-filterbank coefficients of 11 successive frames. The acoustic model is a 5-layer fully connected network with 1024 neurons at each hidden layer and 3440 output units that produce a distribution over senones. We use ``ReLU'' activation function in intermediate layers and ``softmax'' in the output layer that generates the probabilities of senones. We initialize all weights using the ``he-normal'' initialization technique~\cite{he2015delving}. The loss function for training the baseline model and also adaptation (i.e., $J_{cross}(\theta)$ in our equations) is the cross-entropy between the forced aligned senone labels and the model outputs.

\textbf{Model training} -- In all experiments, we use Adam optimizer to train or adapt the DNN models~\cite{kingma2014adam}. For training the original model with Libri, we use a learning rate (LR) of 0.001; however, we found that smaller learning rates perform better for adapting the original model. Our initial experiments showed that the learning rate of 0.0001 is an effective choice for the model expansion experiments. To train the original model, we employ the early-stopping technique to deal with the over-training problem. Early-stopping is performed by monitoring the performance of the model on a held-out validation set~\cite{khorram2018priori, khorram2019jointly}. However, applying early-stopping is not efficient for the domain expansion task \cite{lwf}; it is because the data of the original domain is not available and performing the early-stopping only on the data of the new domain is just beneficial for the new domain, and it may significantly reduce the model performance on the original domain \cite{lwf}. In continual learning, a common approach is to perform a fixed number of iterations to train the new model~\cite{kemker2018measuring}. In this study, we found that fine-tuning the original model converges to an optimum solution in 20 epochs. To investigate the efficacy of each approach, we evaluate their performance in three independent scenarios that consider IND, AUS, and HIS accents as the new domains.


\vspace{-0.5em}
\subsection{Results}

We conduct several experiments to evaluate the performance of the domain expansion methods explored in this study. As mention in previous sections, each domain expansion technique has a controlling hyper-parameter that provides a compromise between keeping the model performance for the original data and learning the new domains. In the first set of experiments, we tune these hyper-parameters to achieve the best overall performance for both the original and the new domains. The results for all approaches are summarized in Table \ref{table:1}. We also report the results of multi-condition training \cite{seltzer2013investigation} in which the model is trained with both original and new domains by pooling their data. The performance of the multi-condition system can be considered as an upper bound for the performance of domain-expansion methods. 

In figure \ref{fig:compare}, we study the effect of changing the above-mentioned hyper-parameters for three model expansion techniques: WCA, EWC and SKLD. This figure shows which method is better in proving a trade-off between retaining the original model and learning the new domain.



\textbf{Forgetting effect}. The original model performs well for the Libri clean test set that matches the training set conditions; however, for the unseen domains, the performance of the model degrades significantly. Fine-tuning (FT) this model to the unseen domains results in a significant improvement in WER for the new domains, but the model performance for the original domain drops dramatically (Table \ref{table:1}). Figure \ref{fig:Forgetting} shows the rate of forgetting the information of the original data as we fine-tune the model to AUS accent (FT-Original). We also show how the SKLD approach performs in the same setting. SKLD can successfully preserve the model performance for the original data while learning the new data. The overall performance of the model on both old and new domains demonstrates that SKLD performs significantly better than naive fine-tuning for domain expansion. 

\textbf{The performance of WCA and EWC}. 
WCA as a naive domain expansion method performs significantly better than fine-tuning in finding a compromise between the performance of the original and new domains. For EWC experiments, since the diagonal of the Fisher information matrix F is zero for many of the original model's weights, simply applying EWC does not preserve the model performance. We found that adding an empirically determined value of 1 to the elements of the matrix addresses the problem. EWC outperforms WCA in all the experiments (Table \ref{table:1} and Figure \ref{fig:compare}), which demonstrates the efficacy of the Fisher information matrix in preserving the learned information of the original data. For example, for IND accent in Table \ref{table:1}, both approaches achieve 17.6 WER for the new data, while EWC achieves a relative WER improvement of +3.8\% vs. WCA. 

\textbf{The performance of SKLD and Hybrid SKLD-EWC}. 
SKLD significantly outperforms all other single domain-expansion approaches yielding a relative WER improvement of +8.8\% and +7.6\% vs. EWC for AUS and IND accent, respectively (Table \ref{table:1}). For HIS accent, SKLD is still slightly better than EWC. The hybrid SKLD-EWC that benefits from both SKLD and EWC results in the best overall performance. Comparing the performance of domain expansion techniques with multi-condition training in Table \ref{table:1} indicates that we have achieved comparable performance with multi-condition training which uses the original training data we consider unavailable for the domain expansion approaches.

\section{Conclusions}
In this paper, we explore several continual learning-based domain expansion techniques as an effective solution for domain mismatch problem in ASR systems. We examine the efficacy of the approaches through experiments on adapting a model trained with native English to three different English accents: Australian, Hispanic and Indian. We demonstrate that simply adapting the original model to the target domains results in a significant performance degradation of the adapted model for the original data. However, we demonstrate that SKLD and hybrid SKLD-EWC are effective in adapting the native English model to the new accents while retaining the performance of the adapted model for native English. The proposed SKLD-EWC outperformed other existing approaches such as fine-tuning, WCA, and EWC. 



\bibliographystyle{IEEEbib}
\bibliography{strings,refs}

\end{document}